\documentclass[aps,prl,twocolumn,showpacs,amsmath,amssymb,superscriptaddress]{revtex4}
\pdfoutput=1
\usepackage{graphicx}
\usepackage{amssymb}
\usepackage{natbib}
\usepackage{calc}
\usepackage{SIunits}
\usepackage{textcomp}

\begin{document}

\title{Spin-Excitations Anisotropy in the Bilayer Iron-Based Superconductor CaKFe$_4$As$_4$}

\author{Tao Xie}
\affiliation{Beijing National Laboratory for Condensed Matter
Physics, Institute of Physics, Chinese Academy of Sciences, Beijing 100190, China}
\affiliation{University of Chinese Academy of Sciences, Beijing 100049, China}
\author{Chang Liu}
\affiliation{Beijing National Laboratory for Condensed Matter
Physics, Institute of Physics, Chinese Academy of Sciences, Beijing 100190, China}
\affiliation{University of Chinese Academy of Sciences, Beijing 100049, China}
\author{Fr\'{e}d\'{e}ric Bourdarot}
\affiliation{Universit\'{e} Grenoble Alpes, CEA, INAC, MEM MDN, F-38000 Grenoble, France}
\author{Louis-Pierre Regnault}
\affiliation{Intitut Laue Langevin, 71 avenue des Martyrs, CS 20156, 38042 Grenoble Cedex, France}
\author{Shiliang Li}
\affiliation{Beijing National Laboratory for Condensed Matter
Physics, Institute of Physics, Chinese Academy of Sciences, Beijing
100190, China}
\affiliation{University of Chinese Academy of Sciences, Beijing 100049, China}
\affiliation{Songshan Lake Materials Laboratory, Dongguan, Guangdong 523808, China}
\author{Huiqian Luo}
\email{hqluo@iphy.ac.cn}
\affiliation{Beijing National Laboratory for Condensed Matter
Physics, Institute of Physics, Chinese Academy of Sciences, Beijing
100190, China}
\affiliation{Songshan Lake Materials Laboratory, Dongguan, Guangdong 523808, China}

\date{\today}

\begin{abstract}
We use polarized inelastic neutron scattering to study the spin-excitations anisotropy in the bilayer iron-based superconductor CaKFe$_4$As$_4$ ($T_c$ = 35 K). In the superconducting state, both odd and even $L-$modulations of spin resonance have been observed in our previous unpolarized neutron scattering experiments (T. Xie {\it et al.} Phys. Rev. Lett. {\bf 120}, 267003 (2018)). Here we find that the high-energy even mode ($\sim 18$ meV) is isotropic in spin space, but the low-energy odd modes consist of a $c-$axis polarized mode around 9 meV along with another partially overlapped in-plane mode around 12 meV. We argue that such spin anisotropy is induced by the spin-orbit coupling in the spin-vortex-type fluctuations of this unique compound. The spin anisotropy is strongly affected by the superconductivity, where it is weak below 6 meV in the normal state and then transferred to higher energy and further enhanced in the odd mode of spin resonance below $T_c$.

\end{abstract}

\maketitle
The neutron spin resonance, a collective mode of spin excitations appears at a specific energy around the antiferromagnetic (AF) wave vector of the parent compounds in the superconducting state, is generally believed as a curial evidence for the spin-fluctuations mediated superconducting pairing in unconventional superconductors \cite{Chen2014,Tranquada2014,Dai2015,Inosov2016,Meschrig2006,gyu2009,pdjohnson}. Theoretically, the spin resonance mode is argued to be a collective bosonic mode from singlet-triplet excitations of the Cooper pairs, thus it is expected to be isotropic in the spin space due to its spin-1 nature \cite{Meschrig2006}. However, the case in iron-based superconductors is much more complicated. Considering the fact that the spin excitations in these materials have both local-moment and itinerant characteristics\cite{Chen2014,Dai2015,Inosov2016}, either by approximating to the parent compound with strong single-ion anisotropy \cite{Braden2004,Wang2013,Song2018}, or being affected by the orbital ordering via spin-orbit coupling (SOC)  \cite{Johnson2015,Borisenko2016,Yi2017}, a spin anisotropy can be present in the spin resonance mode. While the spin anisotropy commonly exists in the low-energy spin excitations even persists into high temperature well above the N\'{e}el temperature $T_N$, the anisotropy of the spin resonance is different among the iron-based superconductors \cite{Lipscombe2010,Babkevich2011,Liu2012,Zhang2013,Steffens2013,Luo2013,Qureshi2014a,Qureshi2014b,Zhang2014,Song2016,Ma2017,Hu2017,Xie20181}. For example, the spin resonance mode in the 11-type compounds FeSe seems to be purely $c-$axis polarized \cite{Ma2017}, but in the 112-type compound Ca$_{1-y}$La$_{y}$Fe$_{1-x}$Ni$_{x}$As$_{2}$, it is completely isotropic \cite{Xie20181}. In the 122-systems such as K, Co, Ni or P doped BaFe$_2$As$_2$ compounds, although the spin resonance modes are either $c-$axis polarized or weakly $a-$axis polarized\cite{Lipscombe2010,Liu2012,Steffens2013,Luo2013,Zhang2014,Song2016,Hu2017}, the maximum energies of spin anisotropy in the superconducting state have been found to be linearly scaling with their superconducting critical temperature $T_c$ \cite{Song2016,Hu2017}.

To further understand the spin anisotropy of the spin resonance and the role of the SOC in iron-based superconductivity, we have performed polarized neutron scattering experiments on a stoichiometric iron-based superconductor CaKFe$_4$As$_4$ with $T_c$ = 35 K [Fig. 1(a)] \cite{Iyo2016,Meier2016}. In this compound, neither the tetragonal-orthorhombic structural transition related to the nematic order nor the long-ranged magnetic order with single ion anisotropy exists, and the absence of chemical dopants gives minimum disorder effect from local impurity. Even though, the SOC may also exist in CaKFe$_4$As$_4$, since it may naturally close to a quantum critical point of a hedgehog spin-vortex crystal (SVC) order, which is discovered in those Ni or Co doped compounds and has two perpendicularly ordered moments with equal magnitudes within the $ab-$plane \cite{Meier2018,Kreyssig2018}. Previously, our unpolarized neutron scattering experiments have revealed triple resonance modes around 9.5 meV, 13 meV and 18.3 meV [Figs. 1(d) and 1(e)]\cite{Xie20182}. All of them show strong $L$-modulations which can be classified into odd and even symmetries with respect to the Fe-Fe distance within one Fe-As bilayer unit[Figs. 1(c)-(e)] \cite{Xie20182,Fong2000,Paihes2004,Sidis2007}. Such separated multiple resonance modes from non-degenerate spin excitations offer a unique chance to study the spin anisotropy of each mode. Here by using polarized neutron scattering analysis [Fig. 1(b)], we have found that the high-energy even mode of spin resonance is isotropic, while the low-energy odd modes become pronouncedly anisotropic with a $c-$axis polarized mode around 9 meV overlapped with a weakly anisotropic in-plane mode around 12 meV. The maximum energy of spin anisotropy also follows the linear scaling with $T_c$ as shown in doped BaFe$_2$As$_2$ systems [Fig. 1(f)]. The spin anisotropy in CaKFe$_4$As$_4$ can be attributed to the SOC under SVC-type fluctuations. It is pushed to higher energy and then strongly enhanced in the spin resonance after cooling down below $T_c$.

\begin{figure}[t]
\includegraphics[width=0.48\textwidth]{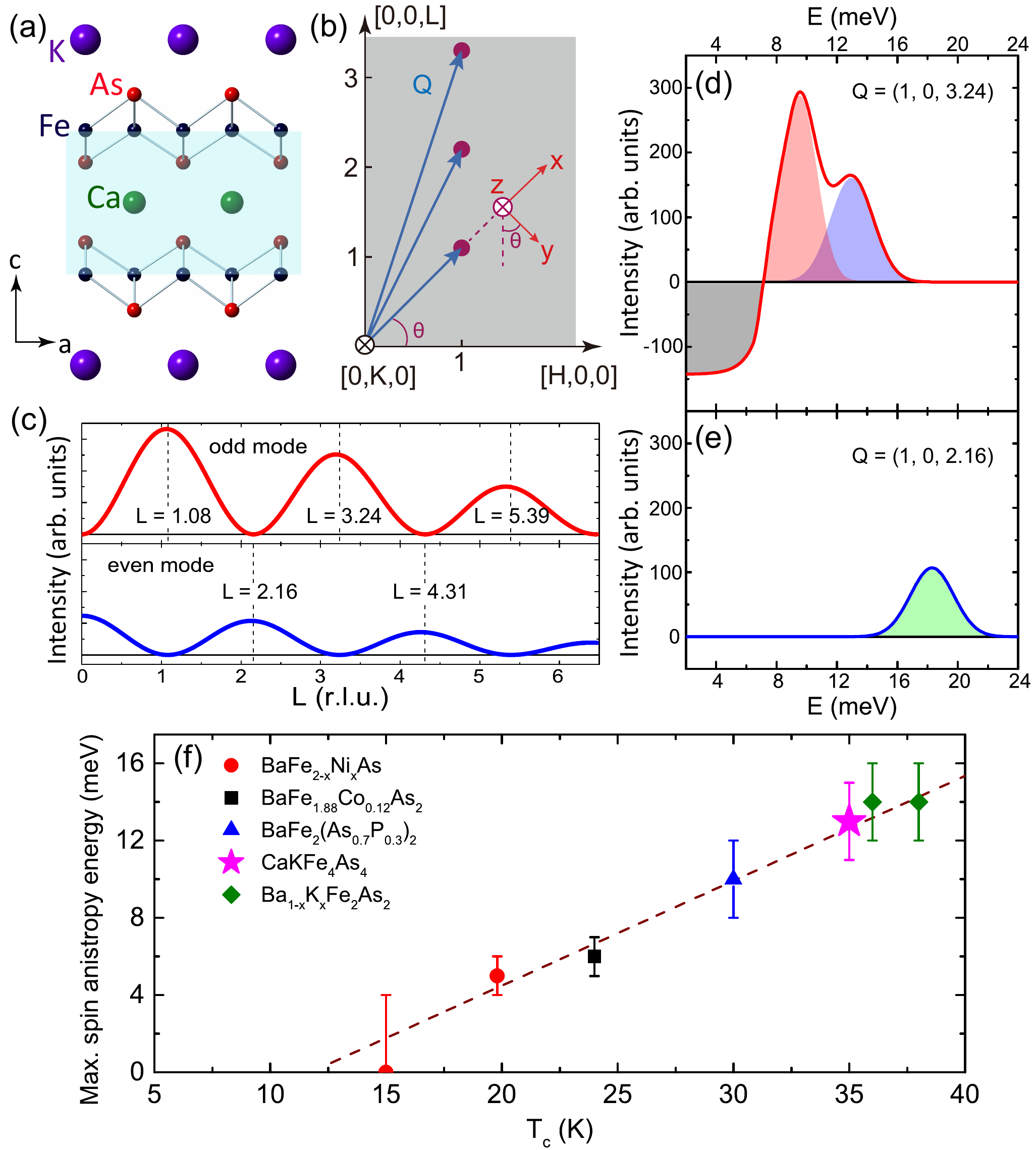}
\caption{
(a) Crystal structure of CaKFe$_4$As$_4$, the lattice axes are indicated in the tetragonal notation, the blue shadow area indicates the Fe-As bilayer unit. (b)The scattering plane and the definition of the spin polarization directions in the reciprocal space. (c) Odd and even $L$-symmetries of the spin resonance. The dashed lines indicate the non-integral $L$ positions with maximum intensities. (d), (e) Schematic diagrams of three spin resonance modes at ${\bf Q} = (1, 0, 3.24)$ (odd modes) and ${\bf Q} = (1, 0, 2.16)$ (even mode) deduced from the intensity differences between $T=1.5$ K and 40 K \cite{Xie20182}.  (f) Linear scaling between the maximum spin anisotropy energy and $T_c$.
}
\end{figure}

\begin{figure}[t]
\includegraphics[width=0.48\textwidth]{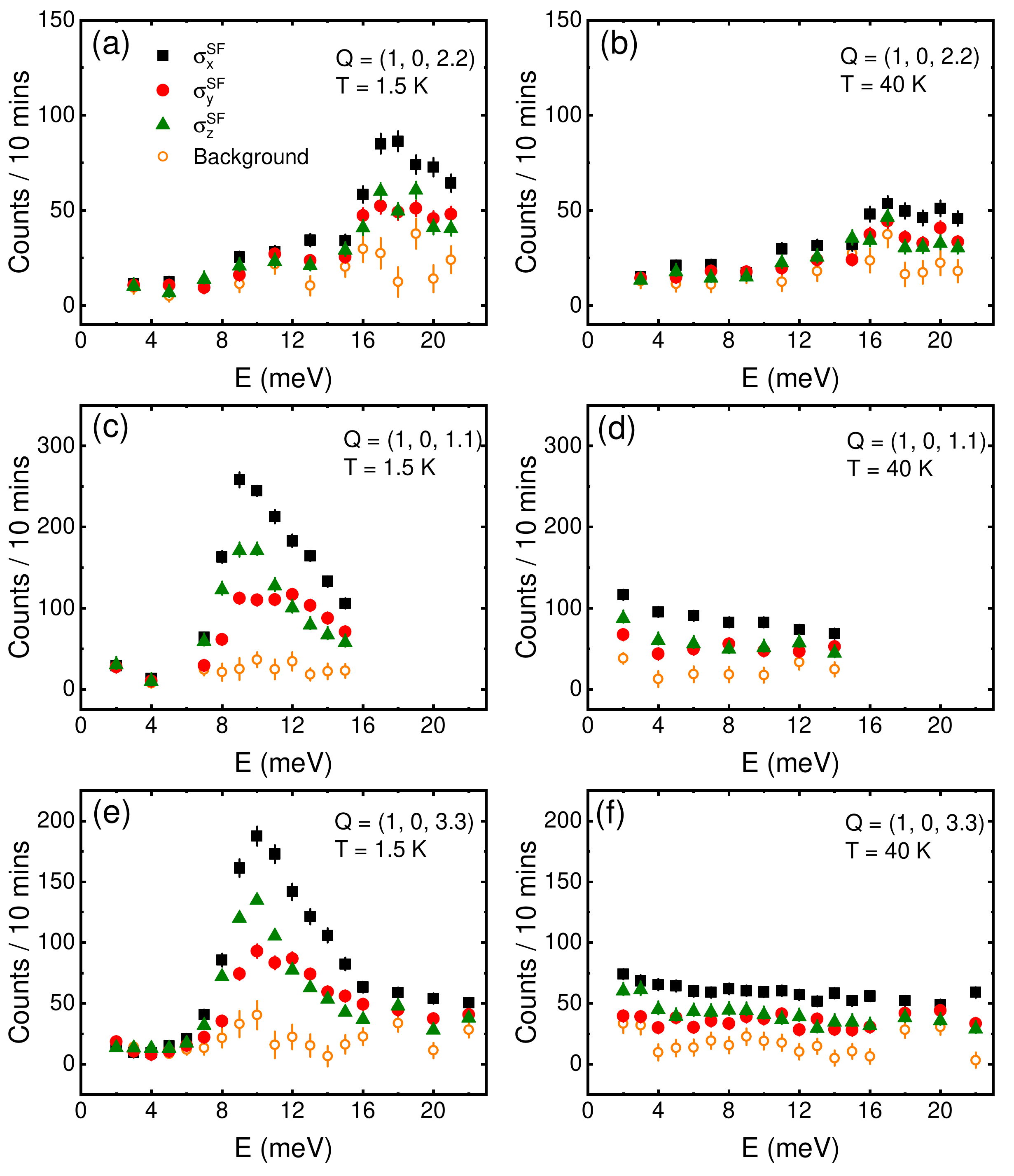}
\caption{ Raw data of constant-${\bf Q}$ scans of $\sigma_{x}^{\textrm{SF}}$, $\sigma_{y}^{\textrm{SF}}$, and $\sigma_{z}^{\textrm{SF}}$ at ${\bf Q}=(1, 0, L)$ with (a) $L = 2.2$, (c) $L = 1.1$, (e) $L = 3.3$ at 1.5 K (superconducting state) and (b) $L = 2.2$, (d) $L = 1.1$, (f) $L = 3.3$ at 40 K (normal state). Open circles are the background estimated by using $B=\sigma_y^{\textrm{SF}}+\sigma_z^{\textrm{SF}}-\sigma_x^{\textrm{SF}}$.
 }
 \end{figure}

We used the same set of CaKFe$_4$As$_4$ single crystals in our previous unpolarized neutron scattering experiments \cite{Xie20182}. The polarized neutron scattering experiments were carried out using the CryoPAD system at the CEA-CRG IN22 thermal triple-axis spectrometer in Institut Laue-Langevin, Grenoble, France, with a fixed final energy $E_f=$ 14.7 meV. The CryoPAD system provides a strictly zero magnetic field environment for the measured sample, thus avoiding errors due to flux inclusion and field expulsion in the superconducting state of the sample \cite{Berna2005}. The scattering plane is $[H, 0, 0] \times [0, 0, L]$ defined using the magnetic unit cell same as before \cite{Xie20182}: $a_M= b_M= 5.45$ \AA, $c=12.63$ \AA, in which the wave vector ${\bf Q}$ at ($q_x$, $q_y$, $q_z$) is $(H,K,L) = (q_xa_M/2\pi, q_yb_M/2\pi, q_zc/2\pi)$ reciprocal lattice units (r.l.u.). We defined the neutron polarization directions as $x, y, z$, with $x$ parallel to $\textbf{Q}$, and $y$ (in scattering plane) and $z$ (out of scattering plane) perpendicular to \textbf{Q} as shown in Fig. 1(b). Since the spin of a neutron has a tendency to flip (spin-flip: SF) after a scattering event with the magnetic excitations from sample. The three neutron SF scattering cross sections can be written as $\sigma_{x}^{\textrm{SF}}$, $\sigma_{y}^{\textrm{SF}}$, and $\sigma_{z}^{\textrm{SF}}$. Because neutron SF scattering is only sensitive to the magnetic excitations/moments that are perpendicular to the momentum transfer ${\bf Q}$ and the neutron spin polarization directions \cite{Lipscombe2010,Babkevich2011,Liu2012}, in our geometry:
\begin{equation}
\begin{array}{l}
\sigma_x^{\textrm{SF}}=cM_y+cM_z+B, \\
\\[1pt]
\sigma_y^{\textrm{SF}}=cM_z+B, \\
\\[1pt]
\sigma_z^{\textrm{SF}}=cM_y+B. \\
\end{array}
\end{equation}
Here, $M_y$ and $M_z$ are the magnitudes of magnetic excitations along the $y$ and $z$ directions, $B$ is the constant background, and $c=(R-1)/(R+1)$. The spin flipping ratio $R$ represents the quality of the neutron spin polarization, defined by the leakage of nuclear Bragg peaks(should be non-spin-flip: NSF) into the SF channel: $R$ = $\sigma_{nuclear}^{\textrm{NSF}}$/$\sigma_{nuclear}^{\textrm{SF}}$, which is about 13 in our experiments.
From Eq.(1), the background can be estimated by comparing the SF scattering between all three channels, which is $B=\sigma_y^{\textrm{SF}}+\sigma_z^{\textrm{SF}}-\sigma_x^{\textrm{SF}}$. $M_y$ and $M_z$ in Eq. (1) can be further written as the combinations of the magnetic excitations along the lattice axes $a_M, b_M$ and $c$:
\begin{equation}
\begin{array}{l}
M_y=M_a\sin^2\theta+M_c\cos^2\theta, \\
\\[1pt]
M_z=M_b, \\
\end{array}
\end{equation}
where $\theta$ is the angle between $\textbf{Q}$ and the $(H, 0, 0)$ direction [Fig. 1(b)]. By measuring the SF scattering cross sections at least two unparallel \textbf{Q} positions with same energy transfer, we can deduce all three components $M_a$, $M_b$ and $M_c$ \cite{Supplementary}. In this method, we don't have to assume $M_a=M_b$ in the calculation \cite{Wang2013,Luo2013,Hu2017,Song2016,Xie20181}, even it is very likely in this stoichiometric material with tetragonal lattice structure \cite{Iyo2016,Meier2016}.

\begin{figure}[t]
\includegraphics[width=0.48\textwidth]{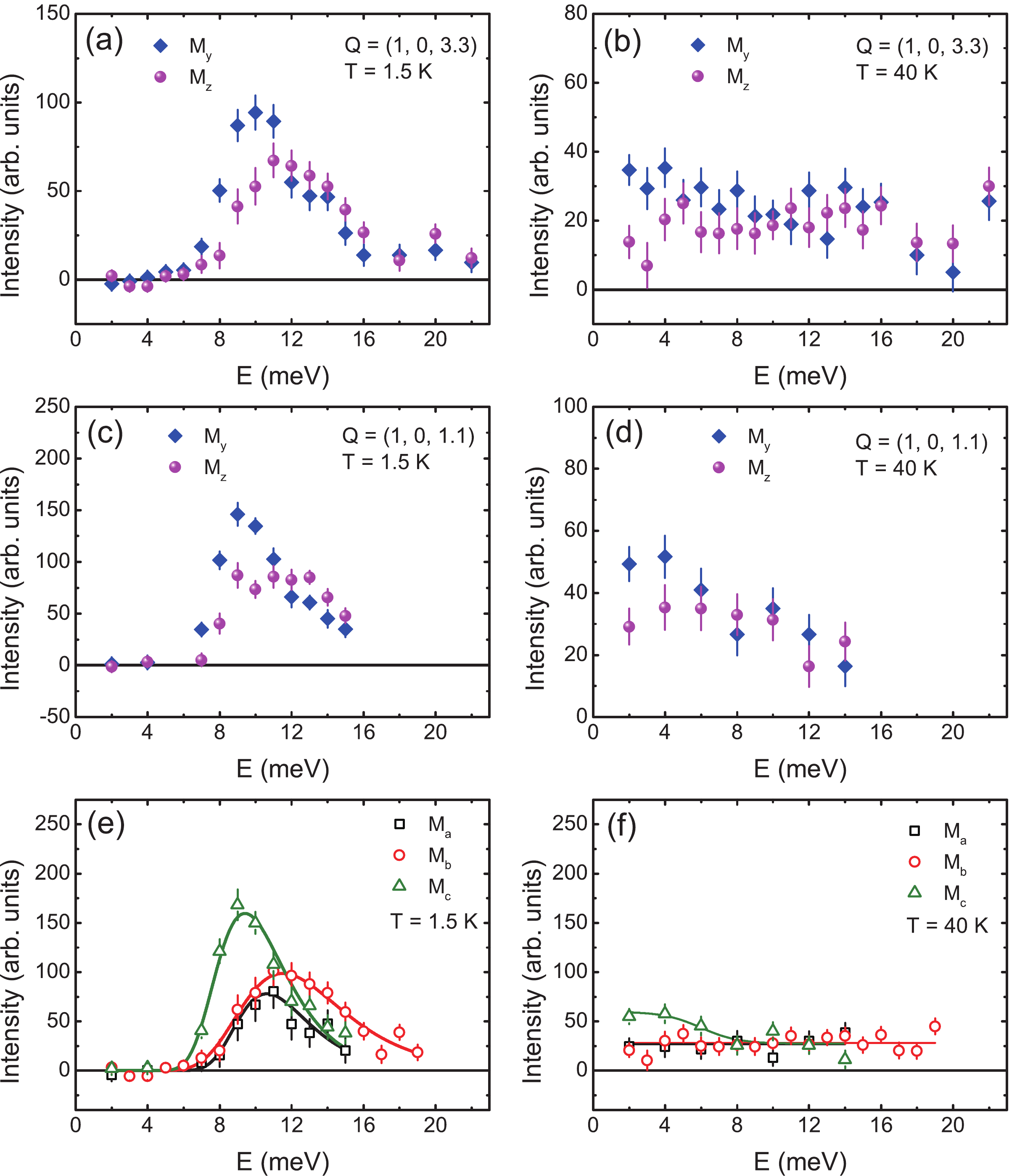}
\caption{ (a), (b) Energy dependence of $M_{y}$ and $M_{z}$ at ${\bf Q}=(1, 0, 3.3)$ below and above $T_c$. (c), (d) Energy dependence of $M_{y}$ and $M_{z}$ at ${\bf Q}=(1, 0, 1.1)$ below and above $T_c$. (e, f) Energy dependence of $M_{a}$, $M_{b}$, and $M_{c}$ below and above $T_c$, where the solid lines are guides to eyes.
 }
\end{figure}

\begin{figure}[t]
\includegraphics[width=0.48\textwidth]{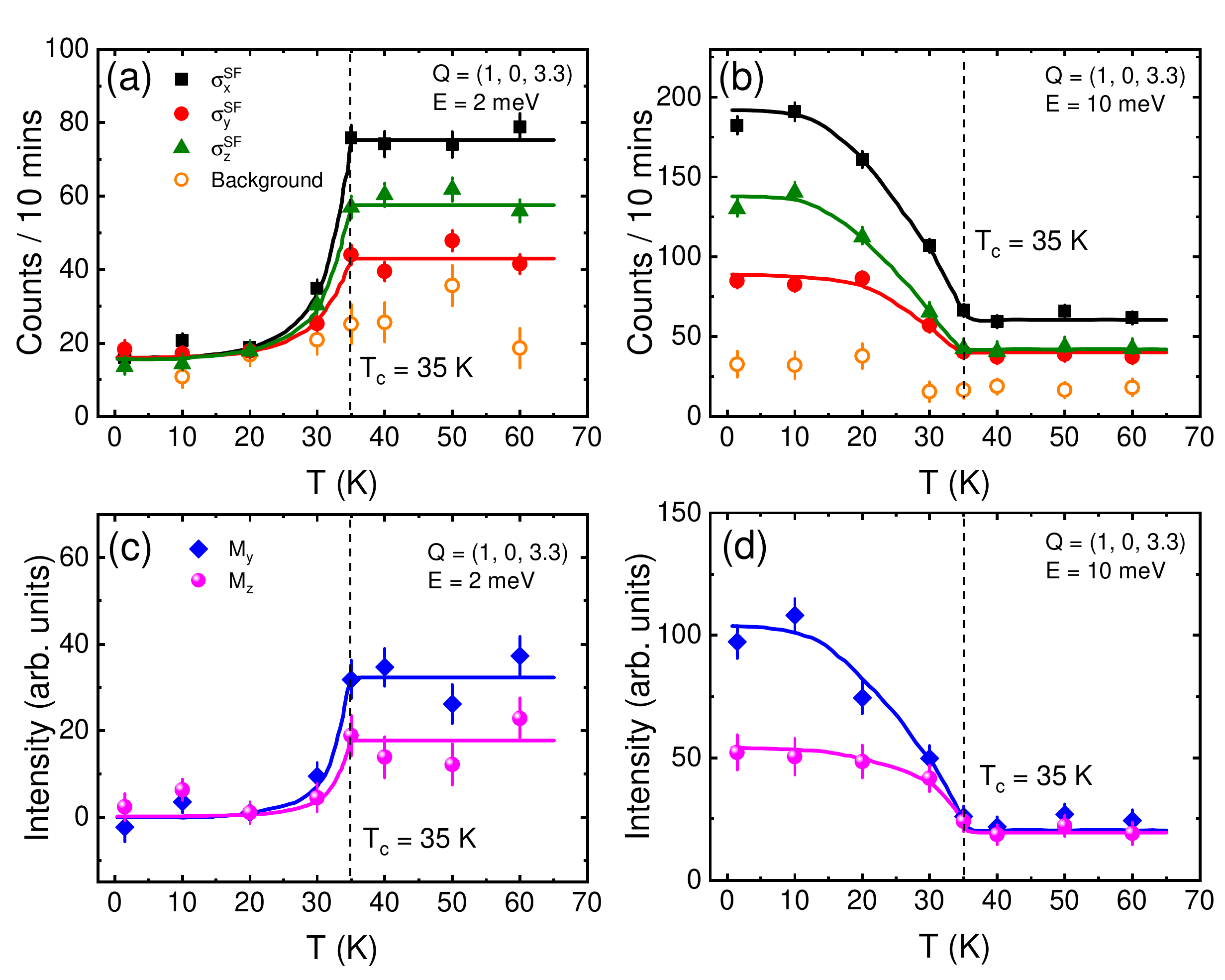}
\caption{
(a), (b) Temperature dependence of the $\sigma_{x}^{\textrm{SF}}$, $\sigma_{y}^{\textrm{SF}}$, and $\sigma_{z}^{\textrm{SF}}$ at ${\bf Q}=(1, 0, 3.3)$ at 2 meV and 10 meV, where the open circles are estimated backgrounds. (c), (d) Temperature dependence of the $M_{y}$ and $M_{z}$ at ${\bf Q}=(1, 0, 3.3)$ at 2 meV and 10 meV. The solid lines are guides to eyes. The vertical dashed lines indicate $T_c$.
 }
\end{figure}

Since the maximum intensities of the spin resonance modes locate at non-integer $L$ positions like $L$  = 1.08, 3.24, 5.39 for the odd modes, and $L$ = 2.16, 4.31, 6.47 for the even mode [Fig. 1(c)]\cite{Xie20182}, we picked the non-integer $L$ indexes $L=$ 1.1, 2.2 and 3.3 to do the measurements for accuracy. Figure 2 gives the raw energy scans of $\sigma_{x}^{\textrm{SF}}$, $\sigma_{y}^{\textrm{SF}}$, and $\sigma_{z}^{\textrm{SF}}$ at ${\bf Q}=(1, 0, L)$. In contrast to the almost featureless results in the normal state [Figs. 2(b), 2(d) and 2(f)], the spin excitations in the superconducting state show clear spin resonance peaks [Figs. 2(a), 2(c) and 2(e)] around 18 meV for $L = 2.2$ (even mode) and 10 meV for $L = 1.1$ and $L = 3.3$ (odd modes) in all the SF channels. These observed spin resonance modes are consistent with the previous unpolarized results \cite{Xie20182}.  If the spin excitations are isotropic in spin space, there will be no difference between $M_y$ and $M_z$, thus $\sigma_y^{\textrm{SF}}=\sigma_z^{\textrm{SF}}=(\sigma_x^{\textrm{SF}}+B)/2$ \cite{Xie20181,Luo2013}. It is indeed the case at ${\bf Q}=(1, 0, 2.2)$ in both the superconducting and normal state [Figs. 2(a) and 2(b)], suggesting the even mode is completely isotropic. While the results for the odd modes at ${\bf Q}=(1, 0, 3.3) $ and $(1, 0, 1.1)$ are very different. In detail, in the superconducting state, the spin excitations are remarkably anisotropic in the energy window $E= 6 \thicksim 15$ meV for the odd modes of spin resonance, and there is a full spin gap below 5 meV due to the identical intensities for all three channels ($\sigma_x^{\textrm{SF}}=\sigma_y^{\textrm{SF}}=\sigma_z^{\textrm{SF}}$), which only measures the isotropic backgrounds [Fig. 2(c) and (e)]. Above $T_c$, the signals become almost isotropic except for a weak anisotropy remained at low energies below 6 meV [Fig. 2(d) and (f)]. The intensity differences between ${\bf Q}=(1, 0, 3.3)$ and $(1, 0, 1.1)$ come from the magnetic form factor and the instrument resolution effects \cite{Supplementary,Luo2013}.

Using Eq. (1), we can obtain the magnitudes of magnetic excitations along $y$ and $z$ polarization directions: $cM_{y}$ = $\sigma_{x}^{\textrm{SF}}$ $-$ $\sigma_{y}^{\textrm{SF}}$ and $cM_{z}$ = $\sigma_{x}^{\textrm{SF}}$ $-$ $\sigma_{z}^{\textrm{SF}}$. The intensity difference of spin excitations along the $y$ and $z$ directions displays the anisotropic behavior more visually  [Figs. 3(a)-3(d)].
With $M_y$ and $M_z$ derived from two equivalent ${\bf Q}$ positions ${\bf Q}=(1, 0, 1.1)$ and ${\bf Q}=(1, 0, 3.3)$, we can further calculate $M_a$, $M_b$ and $M_c$ from Eq. (2). The angle $\theta$ between $\textbf{Q}$ and $\textbf{a}$$^{\ast}$ for $L$ = 1.1 and 3.3 are 25.5$\degree$ and 55$\degree$, respectively \cite{Supplementary}. Figures 3(e) and 3(f) display the energy dependence of $M_{a}$, $M_{b}$, and $M_{c}$ in the superconducting state and normal state, respectively. We can identify a $c$-axis polarized resonance mode around 9 meV overlapped with another weakly anisotropic in-plane mode around 12 meV, corresponding to the two overlapped odd modes in the previous unpolarized report \cite{Xie20182}. The spin gap of all channels seems to develop at the same energy $E=6$ meV. The weak anisotropy in the normal state at low energy is also partially $c$-axis polarized below 6 meV, as shown in Fig. 3(f).

To track the anisotropy of spin gap and spin resonance induced by the superconducting transition in CaKFe$_4$As$_4$, we have performed temperature scans at ${\bf Q}=(1, 0, 3.3)$ with energy transfer $E$ = 2 meV and 10 meV, the results are shown in Fig. 4. At $E$ = 2 meV, the spin anisotropy of $\sigma_{z}^{\textrm{SF}}$ $>$ $\sigma_{y}^{\textrm{SF}}$ ($M_y > M_z$) can be clearly seen in the normal state and gradually become invisible below $T_c$ for the zero intensity of the full spin gap opening in all the SF channels [Figs. 4(a) and (c)]. For $E$ = 10 meV, the results are completely opposite: the same type of spin anisotropy ($M_y > M_z$) develops only below $T_c$ when entering into the superconducting state [Figs. 4(b) and (d)]. Such intimate relationships between the spin anisotropy and the superconducting transition indicates that the spin anisotropy is strongly affected by the superconductivity in CaKFe$_4$As$_4$.

The spin anisotropy in CaKFe$_4$As$_4$ gives us new clues to understand the SOC in the superconducting state. Firstly, the spin anisotropy solely presented in the odd modes of spin resonance most likely originates from the SOC. Since the odd and even modes of spin resonance are corresponding to the symmetric and antisymmetric states of the non-degenerate magnetic excitations in FeAs inter-bilayer \cite{Xie20182,Fong2000,Paihes2004,Sidis2007}, both of them should be isotropic in spin-space, assuming that the spin resonance is indeed from a spin-1 exciton of the singlet Cooper pairs and there is no SOC. However, with sufficient SOC, the orbital-selective electron correlations and superconducting pairings would possibly introduce spin-excitations anisotropy \cite{Johnson2015,Borisenko2016,Yi2017}. Secondly, the spin anisotropy from SOC is limited at low-energy spin excitations both in the normal state and the superconducting state, thus the high-energy even mode of spin resonance can keep isotropic. In fact, for many iron-based superconductors explored so far, the anisotropy of spin resonance mostly exists at the low energy part of the resonance peak, while above $E_R$ it is nearly isotropic \cite{Steffens2013,Luo2013,Qureshi2014a,Qureshi2014b,Zhang2014,Song2016,Ma2017,Hu2017,Zhang2013,Babkevich2011}. In the FeSe system with sufficiently low $T_c$ (about 8 K) and $E_R$ (about 4 meV), the spin resonance peak can be fully $c-$axis polarized \cite{Ma2017}. Thirdly, the SOC is significantly enhanced in the superconducting state. The superconducting gap opening below $T_c$, not only evokes the collective spin resonance modes, but also partially even fully gaps the low-energy spin excitations. This effect transfers the small spin anisotropy in the normal state to higher energy, which is boosted by the spin resonance in the superconducting state. Because the resonance energy is proportional to the superconducting gaps and $T_c$\cite{Xie20181,Xie20182}, a positive relation between the maximum energy of spin anisotropy and $T_c$  would be expected in those iron-based superconductors with similar strength of SOC [Fig.1(f)] \cite{Song2016,Hu2017}.

The character of SOC is an important key to understand the magnetic structure and superconductivity in the iron-based superconductors \cite{Ma2017,Fernandes2019,YLi2019}. In fact, the SOC induced axis-polarized intensity of the spin resonance mode is usually attributed to a proximity to a possible AF instability or structural transition \cite{Steffens2013,Luo2013,Qureshi2014a,Qureshi2014b,Zhang2014,Song2016,Ma2017,Hu2017,Zhang2013,Babkevich2011}. Although the stoichiometric CaKFe$_4$As$_4$ is naturally paramagnetic, a $C_4$ symmetric SVC phase with ordered moments within $ab$-plane emerges after doping 3.3\% Ni or 7.2\% Co to suppress the superconductivity below 20 K, where two wave-vectors have equal magnitudes in a orthogonal geometry ($\langle|\mathbf{M_1}|\rangle=\langle|\mathbf{M_2}|\rangle\neq0$,$\langle\mathbf{M_1}\rangle\cdot\langle\mathbf{M_2}\rangle=0$) \cite{Meier2018,Kreyssig2018}.  Unlike the Ising-nematic order in the stripe-type AF structure, the SVC phase has an order parameter ($\varphi=\langle\mathbf{M_1}\times\mathbf{M_2}\rangle$) that breaks the spin-rotational symmetry from $O(3)$ to $O(2)$ and produces chiral spin currents together with strong SOC \cite{Fernandes2016}.
In other words, assuming one of the moment $\mathbf{M_1}$ is fixed, another moment $\mathbf{M_2}$ would prefer to fluctuate in the plane perpendicular to easy-plane ($ab-$plane), giving $c-$axis polarized magnetic excitations ($M_c$) at low energy first. The other two modes within the easy-plane ($M_a$ and $M_b$) should cost more energy, since it requires both $\mathbf{M_1}$ and $\mathbf{M_2}$ to fluctuate simultaneously along their direction or to rotate synchronously within the plane, not to mention the stronger exchange coupling $J_{ab}> J_c$ \cite{Dai2015,Inosov2016,pdjohnson}. Indeed, this is exactly the case in the spin resonance of CaKFe$_4$As$_4$, for the asymmetric intensity between the $c-$axis polarized mode around 9 meV and the in-plane mode around 12 meV, where slight difference between $M_a$ and $M_b$ is also reasonable due to their different origins. Further polarized neutron scattering experiments on the Ni/Co doped CaKFe$_4$As$_4$ in the spin-vortex state will be helpful to clarify this issue.

In summary, our polarized neutron scattering analysis on CaKFe$_4$As$_4$ suggests that the spin anisotropy only emerges in the low-energy odd mode of spin resonance, while the high-energy even mode is demonstrated to be isotropic in spin space. Although the spin anisotropy is weak at normal state, it can be significantly enhanced by the superconductivity below $T_c$. The $c-$axis polarized mode around 9 meV along with the overlapped in-plane mode around 12 meV, are probably induced by the SOC in the SVC-type fluctuations.

This work is supported by the National Key Research and Development Program of China (2018YFA0704201, 2017YFA0303103, 2017YFA0302903 and 2016YFA0300502), the National Natural Science Foundation of China (11822411, 11961160699, 11674406 and 11674372), the Strategic Priority Research Program (B) of the Chinese Academy of Sciences (CAS) (XDB25000000 and XDB07020300). H. L. is grateful for the support from the Youth Innovation Promotion Association of CAS (2016004) and the Beijing Natural Science Foundation (JQ19002). This work is based on the experiment(4-02-532) performed at IN22 thermal triple-axis neutron scattering spectrometer in the Intitut Laue Langevin (ILL), Grenoble, France, raw data are available in the DOI system (10.5291/ILL-DATA.4-02-532) \cite{RawData}.

\clearpage
\appendix
\section{Supplementary Materials}

In our experiment, the scattering plane is $[H, 0, 0] \times [0, 0, L]$ defined using the magnetic unit cell similar to the 122-type iron pnictides: $a_M= b_M= 5.45$ \AA, $c=12.63$ \AA, in which the wave vector ${\bf Q}$ at ($q_x$, $q_y$, $q_z$) is $(H,K,L) = (q_xa/2\pi, q_yb/2\pi, q_zc/2\pi)$ reciprocal lattice units (r.l.u.). We defined the neutron polarization directions as $x, y, z$, with $x$ parallel to $\textbf{Q}$, and $y$ (in plane) and $z$ (out of plane) perpendicular to \textbf{Q} \cite{Lipscombe2010}. In our geometry:
\begin{equation}
\begin{array}{l}
\sigma_x^{\textrm{SF}}=cM_y+cM_z+B, \\
\\[1pt]
\sigma_y^{\textrm{SF}}=cM_z+B, \\
\\[1pt]
\sigma_z^{\textrm{SF}}=cM_y+B. \\
\end{array}
\end{equation}
Thus $M_y$ and $M_z$ can be written as:
\begin{equation}
\begin{array}{l}
cM_y=\sigma_x^{\textrm{SF}}-\sigma_y^{\textrm{SF}}, \\
\\[1pt]
cM_z=\sigma_x^{\textrm{SF}}-\sigma_z^{\textrm{SF}}. \\
\\[1pt]
\end{array}
\end{equation}
$M_y$ and $M_z$ are also the combinations of the magnetic excitations along the lattice axes $a_M, b_M$ and $c$ \cite{Luo2013,Wang2013,Hu2017,Song2016}:
\begin{equation}
\begin{array}{l}
M_y=M_a\sin^2\theta+M_c\cos^2\theta, \\
\\[1pt]
M_z=M_b, \\
\end{array}
\end{equation}
where $\theta$ is the angle between $\textbf{Q}$ and the $(H, 0, 0)$ direction. Therefore, we can obtain the intensity of magnetic excitations along the lattice axes $M_a$, $M_b$ and $M_c$ by measuring the SF scattering cross sections at two (or more) unparallel but equivalent \textbf{Q} positions. For the odd modes of spin resonance, we measured the SF scattering cross sections at ${\bf Q}_1=(1, 0, 1.1)$ and ${\bf Q}_2=(1, 0, 3.3)$. At ${\bf Q}=(H, 0, L)$, $\tan$$\theta$ = (2$\pi$$L/c$)/(2$\pi$$H/a$) = $La$/$Hc$, $\theta$ = $\arctan$($La$/$Hc$). Thus, for ${\bf Q}_1=(1, 0, 1.1)$, $\theta$$_1$ $\approx$ 25.5$\degree$, for ${\bf Q}_2=(1, 0, 3.3)$, $\theta$$_2$ $\approx$ 50$\degree$, respectively. Then we have:
\begin{equation}
\begin{array}{l}
M_y({\bf Q}_1)=M_a\sin^2 25.5\degree+M_c\cos^2 25.5\degree, \\
\\[1pt]
M_y({\bf Q}_2)=M_a\sin^2 50\degree+M_c\cos^2 50\degree. \\
\end{array}
\end{equation}
In principle, $M_z$ is fully identical to $M_b$. However, the experimental results will give $M_z({\bf Q}_1)$$\neq$$M_z({\bf Q}_2)$ for the normalization effect from the magnetic form factors and instrumental resolution at different ${\bf Q}_1$ and ${\bf Q}_2$. Here we simply use a factor $e$ to represent such effects and suppose it is temperature independent. It can be determined by the ratio of $M_z({\bf Q}_1)/M_z({\bf Q}_2)$,  $M_y$ has to be normalized by the same ratio as well, namely:
\begin{equation}
\begin{array}{l}
\begin{aligned}
\\[1pt]
M_b&= M_z({\bf Q}_1) = eM_z({\bf Q}_2), \\
\\[1pt]
M_y({\bf Q}_1)&=M_a\sin^2 25.5\degree+M_c\cos^2 25.5\degree\\&=0.185M_a + 0.815M_c,\\
\\[1pt]
eM_y({\bf Q}_2)&=M_a\sin^2 50\degree+M_c\cos^2 50\degree\\&=0.672M_a + 0.328M_c. \\
\end{aligned}
\end{array}
\end{equation}
Therefore, by solving the above equations, we can obtain $M_a$, $M_b$ and $M_c$ in arbitrary units.

\end{document}